\documentclass[journal]{IEEEtran}

\usepackage{cite}
\usepackage{amsmath,amssymb,amsfonts}
\usepackage{algorithmic}
\usepackage[dvipsnames]{xcolor}
\usepackage{colortbl}
\usepackage{pgfplots}
\usepackage{verbatim}
\pgfplotsset{width=7cm,compat=1.8}
\usepgfplotslibrary{fillbetween}
\usepackage{xcolor}
\usepackage{tikz}
\usepackage{tikzscale}
\usepackage{pgfplots}
\usetikzlibrary{patterns}
\pgfplotsset{every tick/.style={black,}}
\usepackage{algorithm,algorithmic}
\usepackage{subcaption}
\usepackage{caption}
\usepackage{amsmath}
\usepackage{booktabs}
\usepackage{multirow}
\usepackage{url}
\usepackage{pgfplotstable}
\usepackage[normalem]{ulem}
\usepackage{stackengine}
\usepackage{arydshln}
\usepackage{ulem}

\pgfplotsset{compat=1.11,
        /pgfplots/ybar legend/.style={
        /pgfplots/legend image code/.code={%
        \draw[##1,/tikz/.cd,bar width=8pt,yshift=-0.2em,bar shift=0pt]
                plot coordinates {(0cm,0.8em)};},
},
}
\newcolumntype{g}{>{\columncolor{blue!10}}c}

\usepackage{authblk}

\begin{document}

\title{
Learning from the Best: Active Learning for Wireless Communications
{\footnotesize \textsuperscript{}}
\thanks{*Equal contribution}
}
\author[$\dag$]{Nasim Soltani*}
\author[$\ddag$]{Jifan Zhang*}
\author[$\dag$]{Batool Salehi}
\author[$\S$]{Debashri Roy}
\author[$\nabla$]{Robert Nowak}
\author[$\dag$]{Kaushik Chowdhury}
\affil[$\dag$]{Electrical and Computer Engineering Department, Northeastern University}
\affil[$\ddag$]{Computer Science Department, University of Wisconsin-Madison}
\affil[$\S$]{Computer Science and Engineering Department, University of Texas at Arlington}
\affil[$\nabla$]{Electrical and Computer Engineering Department, University of Wisconsin-Madison}

\markboth{Accepted in IEEE Wireless Communications Magazine, January 2024}
{}

\maketitle

\begin{abstract}
Collecting an over-the-air wireless communications training dataset for deep learning-based communication tasks is relatively simple. However, labeling the dataset requires expert involvement and domain knowledge, may involve private intellectual properties, and is often computationally and financially expensive. Active learning is an emerging area of research in machine learning that aims to reduce the labeling overhead without accuracy degradation. Active learning algorithms identify the most critical and informative samples in an unlabeled dataset and label only those samples, instead of the complete set. In this paper, we introduce active learning for deep learning applications in wireless communications, and present its different categories. We present a case study of deep learning-based mmWave beam selection, where labeling is performed by a compute-intensive algorithm based on exhaustive search. We evaluate the performance of different active learning algorithms on a publicly available multi-modal dataset with different modalities including image and LiDAR. Our results show that using an active learning algorithm for class-imbalanced datasets can reduce labeling overhead by up to 50\% for this dataset while maintaining the same accuracy as classical training.
\end{abstract}



\section{Introduction}\label{sec:intro}
Deep Learning has revolutionized the field of wireless communications by offering automated solutions for many physical layer (PHY) applications, ranging from signal detection and classification~\cite{waldo}, to security measures in data coding and device authentication~\cite{rf-fingerprinting}, as well as receiver chain design~\cite{nasim-spinn}. In such solutions, a learning algorithm learns the mapping between inputs and outputs (labels) through getting trained on an ideally diverse and comprehensive \emph{labeled} dataset. The best generalization ability on real-life test signals is achieved if the training dataset also contains in-the-wild-collected and real-life radio frequency (RF) signals.

\noindent \textbf{Labeling an RF dataset}: Collecting wild over-the-air RF datasets can be performed by simply recording signals in bands of interest from WiFi, cellular, or other wireless networks, however, labeling such a dataset is no trivial task. Labeling an RF dataset can be especially cumbersome since obtaining these labels requires a high degree of professional knowledge. In the RF domain, labels are obtained using either compute-intensive and time-consuming signal processing algorithms or hand-engineered features by human experts. Overall, we describe 3 broad situations where labeling RF datasets is challenging:

\begin{figure}[t!!!]
    \centering
    \includegraphics[width=\linewidth]{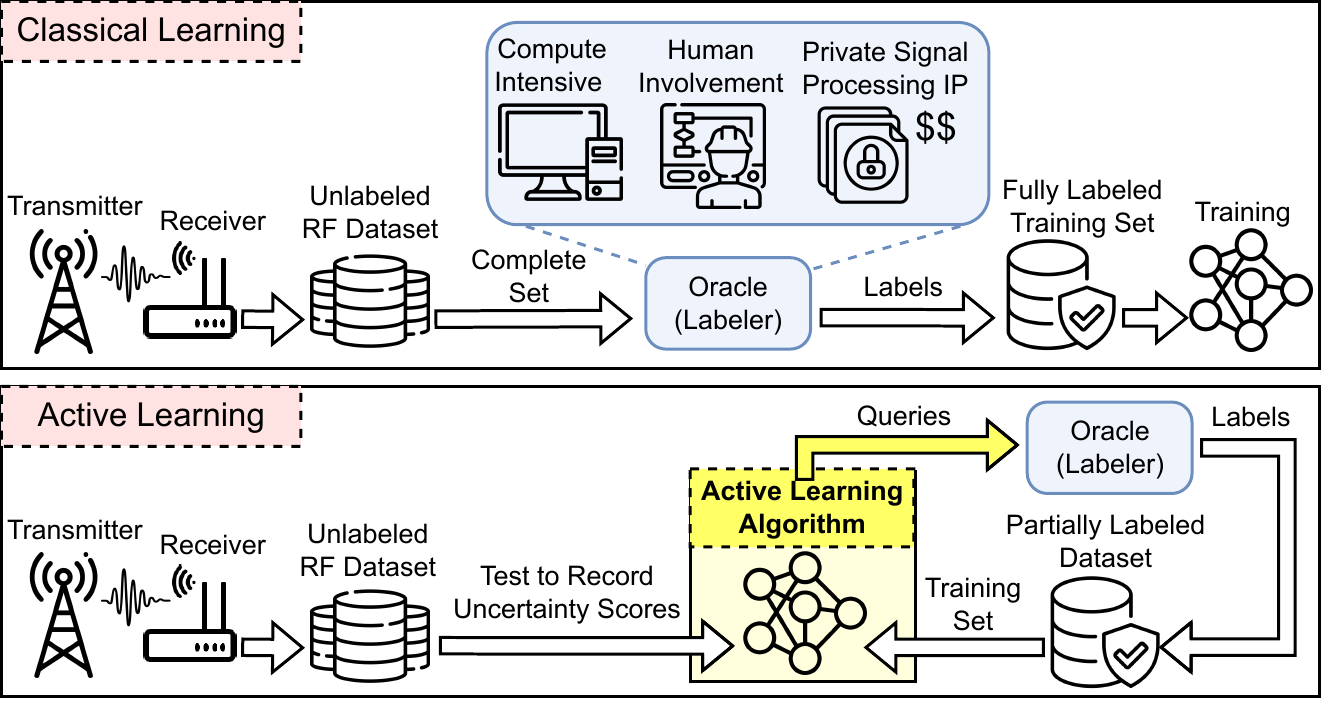}
    \caption{Top: Classical learning where the complete dataset is labeled and used for training. Bottom: Active learning where a subset of the dataset is selected and labeled for training.}
    \label{fig:overview}
\end{figure}

\noindent \textbf{Challenge1. Compute-intensive labeling}: One of the main motivations behind using deep learning instead of the traditional deterministic algorithms in many PHY applications (e.g., RF fingerprinting~\cite{rf-fingerprinting} and modulation classification) is to reduce decision time by substituting the compute-intensive traditional signal processing algorithms with a lighter trained algorithm, in the inference phase. However, the training process still requires the compute-intensive traditional algorithm to provide the labels. Authors in~\cite{icarus} show that (traditional) cyclostationary signal processing (CSP) algorithms require 33.5M floating point operations (FLOPs) to acquire label for one signal in an anomaly detection problem.

\noindent \textbf{Challenge2. Labeling with human in the loop}: Similar to the image processing domain, where human operators label images in the training dataset, many PHY signal processing tasks require human involvement to acquire labels. One such example is CSP, where highly discriminative features are extracted from PHY signals using various periodically time-variant probabilistic parameters. After feature extraction, human involvement is necessary to set specific thresholds for classifying extracted features in a modulation classification problem or detecting the potential presence of a specific waveform~\cite{icarus}.

\noindent \textbf{Challenge3. Financially costly labeling with private intellectual properties (IPs)}: Another challenge is labelling an RF dataset using proprietary and private signal processing IPs designed and owned by telecommunication companies. These IPs are provided to customers for certain end-to-end functions, therefore, their intermediate signals are only internally available and are not provided to the customers through user interfaces. As an example, consider an IP that provides the end-to-end orthogonal frequency division multiplexing (OFDM) receiver processing and data decoding. The input to the IP is the received signal and the output is decoded bits after all the steps of synchronization, channel estimation, equalization, demapping, and decoding~\cite{nasim-spinn}. Since the intermediate outputs such as the estimated channel are not exposed in the interface, if a user wishes to train a channel estimator deep neural network (DNN) using this IP, they have to pay extra for the estimated channel labels. In this case, a method that helps to find the most informative samples to be labeled can reduce the financial labeling cost.

\noindent \textbf{Active Learning to enable ``Learning From the Best"}: Active learning can be used to address the three aforementioned challenges where acquiring signal labels is expensive. In this paper, we introduce \emph{active learning} as a \emph{tool} to train a DNN with reduced number of labeled training samples. As shown in Fig.~\ref{fig:overview}, as opposed to classical training where all the samples in the training dataset are labeled, in active learning the DNN has access to an unlabeled dataset. The active learning algorithm iteratively and adaptively selects the most informative and critical samples from this unlabeled pool using probability scores and queries an information source or {\em oracle} for labels of only those samples. When the oracle provides the labels, a labeled training set is updated iteratively with newly labeled samples and the DNN is trained on the complete labeled training set. The process continues until the desired labeling budget is met. In this way, the learning algorithm learns data distribution using only the most informative samples instead of the whole pool, and hence, \textit{learns from the best}.

In~\cite{yang2018active}, active learning is applied to reduce the expert labeling cost while training random forest models for internet of things (IoT) intrusion detection. However, to the best of our knowledge, \emph{deep} active learning algorithms have not been used in the wireless communications domain before. In this paper, we address a much broader range of concerns around labeling cost of RF datasets. Furthermore, we provide a detailed and first-to-date guide in using different categories of deep active learning algorithms for different PHY applications (Section~\ref{sec:categories}). 
In Section~\ref{sec:mmwave}, we present a case study of active learning for mmWave beam selection using a publicly available multi-modal dataset~\cite{flash} that is extremely class-imbalanced, similar to many other in-the-wild-collected RF datasets. We deploy a properly suited deep active learning algorithm for extremely imbalanced datasets named as GALAXY~\cite{galaxy} on different modalities of the dataset. We show that GALAXY can reduce the labeling need by up to 50\% on this dataset, while maintaining the same accuracy as classical training (Section~\ref{sec:eval}). We present future directions in Section~\ref{sec:future} and conclude the paper in Section~\ref{sec:conclusion}.


\section{Categorizing Active Learning Algorithms for PHY Applications}\label{sec:categories}

\begin{figure*}
    \centering
    \includegraphics[width=\linewidth]{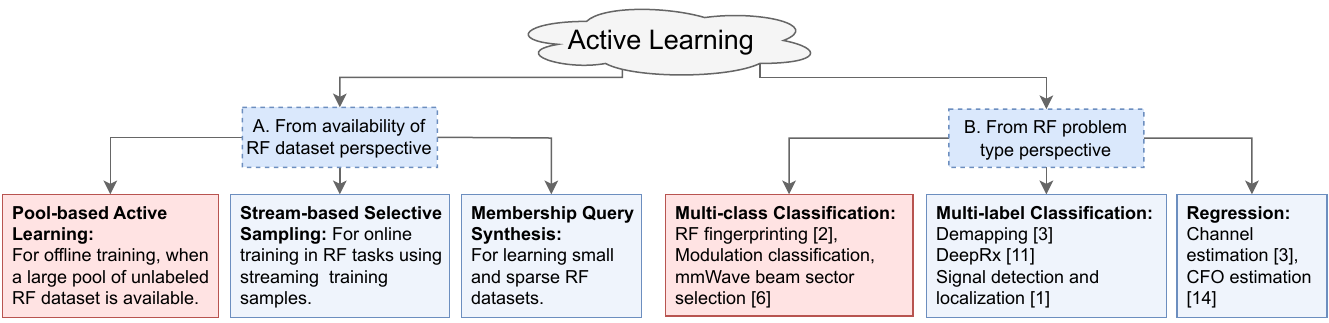}
    \caption{Categorization of active learning algorithms from two different and parallel perspectives of: A. Availability of RF dataset, and B. PHY problem type. In this paper, we study a PHY use case for active learning that falls within the categories marked with red boxes.}
    \label{fig:tree}
\end{figure*}

In this section, we categorize active learning algorithms from two different perspectives: (A) How much of the dataset is available, and (B) What is the deep learning-based PHY problem type. A detailed categorization is described below and a summary is shown in Fig.~\ref{fig:tree}. As it can be seen in Fig.~\ref{fig:tree}, the A and B categories are independent and parallel and any given PHY task can fall in one of the branches under A and another branch under B, depending on the dataset availability and the PHY problem type.

\subsection{Based on Availability of RF Dataset}
\subsubsection{Pool-based Active Learning}
Pool-based active learning is probably the best performing active learning scenario, where the highest performance can be achieved with fewest labeled samples. This high performance is achieved at the expense of a large pool of collected data being available. The pool-based active learning flow is described in the following steps: (i) A specific labeling budget is considered and a batch size, $\mathcal{N}$, is set by the user in the beginning of training. (ii) In the first active learning iteration, $\mathcal{N}$ random samples are drawn from the unlabeled pool and the oracle is queried for the labels of that batch. (iii) The model is trained fully for multiple epochs on the labeled batch (i.e., labeled training set). (iv) In the beginning of next iteration, the so-far trained model is tested on the whole pool and predictions are recorded. (v) A specific active learning algorithm~\cite{holzmuller2022framework} becomes effective to select another $\mathcal{N}$ unlabeled samples --based on the prediction results-- to be queried in the current iteration. (vi) The labels of the new batch are provided by the oracle and the training set is updated with the newly labeled batch. (vii) The model is trained on the current labeled training set, and the steps are repeated starting step (iv) until the labeling  budget is met. 

Pool-based active learning is suitable for offline training tasks, which is ideal for PHY applications too, as it obviates the requirement of specialized hardware for training on edge devices. Offline learning is pragmatic for PHY applications where the discriminating factors of the dataset remain consistent over time, in different heat degrees, in different environments, and under different wireless channels. Examples of such tasks could be signal detection~\cite{waldo}, demapping~\cite{nasim-spinn}, decoding~\cite{nasim-spinn}, etc. In such cases, if a pool of PHY signals is collected, pool-based active learning can be deployed to train an equally robust model with fewer labeled samples.

\subsubsection{Stream-based Selective Sampling} 
In this scenario, the learner receives training samples one by one and decides whether to query or discard them. The key assumption to use this scenario is that obtaining an unlabeled sample is free or inexpensive, which is the case for most PHY applications with either synthetic data generation or over-the-air data collection. As this scenario does not incorporate the assumption of accessing a full dataset, it can be deployed to deep learning-based PHY applications that require online training. Such cases usually encompass applications where a significantly impactful factor changes between the training and deployment phases. In this case, a deep learning algorithm needs to go through a phase of online training as the environment changes. Among the very environment-dependent PHY applications that can benefit from stream-based selective sampling are channel estimation~\cite{nasim-spinn}, and beam selection~\cite{flash} in new environments. In such cases, stream-based selective sampling could be used to skip labeling some of the uninformative samples, and hence reduce queries on the expensive traditional algorithms. This category of active learning algorithms also obviates the need for a large data storage to contain a large pool of unlabeled data. The decision of whether to query or discard a new sample can be taken based on an informativeness measure or by defining an explicit region of uncertainty. 

\subsubsection{Membership Query Synthesis}
In this scenario, the learner creates new training samples in the input space. These new samples could be generated by creating random inputs in the input space or by augmenting the existing training samples. After sample generation, the learner queries the oracle to provide the label for the newly generated sample. Membership query synthesis is specially helpful if the training dataset is small and sparse, however, labeling some of the generated samples could be awkward for the oracle. As an example, consider deploying membership query synthesis to train a modulation classifier DNN on a small and sparse training set. In this case, the learner generates new samples in the input space and queries a classical signal processing algorithm (i.e., an oracle) for the label (a.k.a., modulation class). However, since the sample is synthetically generated and is not actually modulated by a certain scheme, it might be labeled as noise by the oracle.

\subsection{Based on PHY Problem Type}

\subsubsection{Multi-class Classification Problems}\label{sec:multi-class}
Multi-class classification problems are tasks where each input is predicted to be a member of \emph{one} specific class. Examples of such PHY applications are device authentication (a.k.a., RF fingerprinting)~\cite{rf-fingerprinting}, modulation classification, waveform (protocol) classification, and mmWave beam selection~\cite{flash} that is the case study in this paper (see Section~\ref{sec:mmwave}). In machine learning literature, authors in~\cite{ash2021gone} utilize a hybrid of uncertainty and diversity-based strategies for image, tabular, and language class-balanced datasets, which generally performs well against other existing algorithms. Authors in~\cite{zhang2023labelbench} study active learning for large models and datasets through extensive experiments, and demonstrate as the model and dataset sizes increase, label-efficiency gain also increases and the benefits from active learning are highlighted. Additionally, authors in~\cite{galaxy} study the extreme class-imbalanced settings and significantly reduce the labeling cost through balancing the collected labels, while choosing the most uncertain samples in benchmark datasets such as CIFAR.

\subsubsection{Multi-label Classification Problems}
Multi-label classification problems are tasks where each input is a member of \emph{multiple} classes instead of just one. A demapper DNN that converts symbols to bits is considered a multi-label classifier as several output bits can be 1 for each input symbol~\cite{nasim-spinn}. The same output type is designed in~\cite{deeprx}, which proposes a DNN-based 5G receiver. Multi-label classification is previously used also for detecting multiple waveforms in a spectrogram~\cite{waldo} through YOLOv3 framework. In machine learning, there are two types of multi-label queries: sample-based and sample-label-based. Sample-based annotation provides all associated labels of a sample at a time, whereas sample-label-based annotation only gives the binary association between a sample and a particular label. Authors in~\cite{zhang2023algorithm} study the class-imbalanced settings with sample-based annotation by balancing number of labels in each class. In addition, authors in~\cite{citovsky2021batch} develop a strategy for sample-label-based annotation.

\subsubsection{Regression Problems}
Regression PHY problems are applications where a single value or a vector of values are predicted for each input signal. Examples of such problems are carrier frequency offset (CFO) estimation~\cite{pronto} and channel estimation~\cite{nasim-spinn}, respectively. In machine learning literature, authors in~\cite{tsymbalov2018dropout} propose an uncertainty-based strategy by querying samples with the highest variance of inference outputs when applying Monte Carlo dropout to a DNN. In order to take advantage of both uncertainty-based and diversity-based concepts, \cite{ash2021gone} proposes an optimal design strategy by utilizing Fisher information.

More details about active learning algorithms can be found in~\cite{holzmuller2022framework} for interested readers. In the rest of this paper, we focus on a pool-based multi-class classification example use case that is shown with red boxes in Fig.~\ref{fig:tree}.


\section{Active Learning for mmWave Beam Selection}\label{sec:mmwave}

In this section, we introduce mmWave beam selection as an example application to show the benefit of active learning for a deep-learning task in the wireless communications domain. We describe the multi-modal mmWave dataset, and describe the active learning algorithm used to learn this dataset in details.

\subsection{Millimeter-Wave Beam Selection}
Millimeter-Wave beam selection is a PHY application, where the input is the set of collected beams and the output is the ``best'' beam index. Traditionally the best beam is selected through a compute-intensive and time-consuming algorithm based on an exhaustive search on all the possible beams. Authors in~\cite{flash} propose using DNNs on multi-modal data for limiting the exhaustive search to a smaller subset of top beams. Their proposed DNN-based scheme reduces the beam selection time by 57\% for mobility scenarios in V2X communication compared to the classical exhaustive search. However, training the DNN prior to the deployment phase requires labels that are achieved through the exhaustive search. Therefore, labeling a mmWave dataset for a beam selection task could benefit from active learning to reduce the need for labeled training samples and overall reduce labeling overhead. Active learning provides this reduction by selecting the most informative unlabeled samples and querying the oracle (i.e., exhaustive beam selection algorithm) only for the labels of those samples. To evaluate active learning, we use the FLASH multi-modal dataset~\cite{flash} and formulate beam selection as a multi-class classification pool-based active learning problem.

\subsection{Dataset Description}\label{sec:mmwave-dataset}
\begin{figure}[t!!!]
\centering
    \begin{tikzpicture}
        \begin{axis}[
            bar width = 0.1cm,
            ybar=0pt,
            height=4.5cm,
            width=0.45\textwidth,
            ymin=0,
            ymax=8000,
            enlarge x limits=0.06,
            ybar legend,
            legend style={at={(0,1), font=\small}, legend cell align={left},
                anchor=north west,legend columns=2},
            ylabel={Population},
            xlabel={Class Index},
            y label style = {align=center,text width=3cm},
            xtick={0,5,10,15,20,25,30,33},
            xticklabels = {0,5,10,15,20,25,30,33},
            ymajorgrids=true,
            ]
        \addplot [black, fill=red!20]
        coordinates{ 
        (0,123)(1,353)(2,43)(3,250)(4,115)(5,22)(6,345)(7,507)(8,20)(9,2404)(10,207)(11,331)(12,753)(13,380)(14,252)(15,3059)(16,1000)(17,557)(18,6882)(19,927)(20,39)(21,1771)(22,1199)(23,276)(24,39)(25,2254)(26,411)(27,79)(28,556)(29,4211)(30,30)(31,594)(32,170)(33,552)
        };
        \end{axis}
    \end{tikzpicture}
    \caption{Population per class across 30711 samples in the dataset shows extreme class-imbalance in the dataset. The smallest class is class 8 with 20 and the largest class is class 18 with 6882 members.}
    \label{fig:population-per-class}
\end{figure}
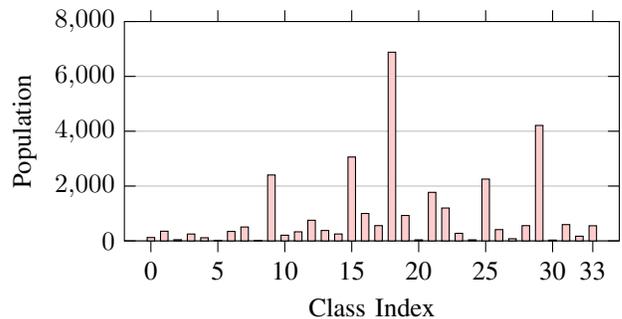

FLASH dataset~\cite{flash} consists of different modalities including camera images and LiDAR collected from an automated car driving in a street, while communicating with a mmWave radio posing as a base station. The mmWave signals are also collected to later be processed and provide the beam indices (i.e., labels) for each location. There are 4 different categories in the dataset consisting of 21 LOS and NLOS scenarios, each consisting of 10 episodes which are 10 different runs of the car in the same scenario. In each location camera image and LiDAR signals are recorded. In the TP-Link Talon AD7200 triband router that is used as the mmWave radio, 34 different beam indices are defined. Therefore, the beam selection problem is formulated as a multi-class classification problem with 34 classes whose indices range from 0 to 33. We count population per class in all the categories, scenarios, and episodes across all the 30711 datapoints in the whole dataset, and demonstrate population per class in Fig.~\ref{fig:population-per-class}. With the smallest class (index 8) with 20 members and the largest (index 18) with 6882 members, we observe an extreme class imbalance in the dataset.

For labeling class-imbalanced datasets such as FLASH, special considerations need to be taken into account so that the samples from larger classes are not favored to be selected over the samples from smaller classes. To incorporate these considerations, authors in~\cite{galaxy} proposes GALAXY.

\subsection{GALAXY Algorithm for Learning Class-imbalanced Datasets}\label{sec:mmwave-galaxy}

GALAXY~\cite{galaxy} is proposed specifically for pool-based active learning in multi-class classification problems on extremely class-imbalanced datasets. \emph{Confidence sampling}, which is a popular uncertainty-based sampling algorithm, selects the samples that show relatively low confidence compared to the rest of samples in the pool, without considering their predicted class. Therefore, in imbalanced datasets confidence sampling is most likely to select samples in larger classes to be labeled. On the other hand, GALAXY finds the optimal decision boundaries through a bisection procedure and selects samples that are both \emph{uncertain} and \emph{class-diverse}.

\begin{figure}[t!!!]
    \centering
    \includegraphics[width=0.7\linewidth]{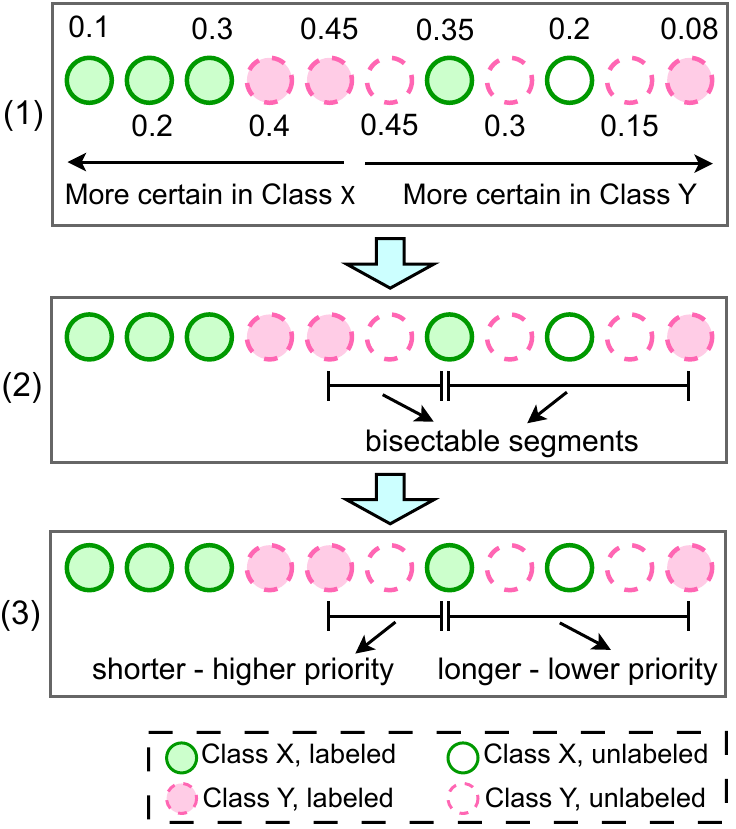}
    \caption{GALAXY algorithm where (1) Uncertainty scores are calculated and the graphs are composed with sorted uncertainty scores for each class X versus all other classes as class Y, (2) Bisectable segments are identified, and (3) 
    Bisectable segments are prioritized and the samples around all identified cuts in the bisectable segments are queried based on the priority.}
    \label{fig:galaxy}
\end{figure}

Similar to all pool-based active learning algorithms, GALAXY comes into action in the beginning of each iteration, after the trained model is tested on the complete pool of samples (see Section~\ref{sec:categories}). GALAXY uses a two-phase procedure. During the first phase, GALAXY calculates a one-versus-all \emph{uncertainty score} for each sample using the predicted probability vector. 
For each example class X, GALAXY sorts the pool samples (including unlabeled and already labeled samples) based on their uncertainty scores and forms a linear graph for each class X with samples as graph nodes, while it considers all other classes as class Y. The samples on the two ends of the graph have the lowest uncertainty scores in classes X and Y. The samples with higher uncertainty are located in between the two end nodes. The goal is to label all pairs of nodes that form an edge and are classified as different classes of X and Y. Such edges are called \emph{cut}s. We call a segment of consecutive nodes on the graph \emph{bisectable}, if it has \emph{labeled} samples of classes X and Y on its two ends, and it contains \emph{no already-labeled} cuts. We note that a bisectable segment always has a cut in it. When we locate a bisectable segment, a bisection procedure is performed, where GALAXY iteratively queries and labels samples one at a time. If there are multiple bisectable segments in each graph, GALAXY prioritizes the shortest segment across \emph{all} graphs to bisect. The second phase starts when there is no bisectable segments left in any graphs. In this phase, GALAXY queries and labels samples around all identified cuts thus far. During this process, if additional bisectable segments appear, GALAXY reverts to the first phase. A simplified overview of GALAXY is shown in Fig.~\ref{fig:galaxy}. 

In each iteration, GALAXY search stops as soon as a batch of $\mathcal{N}$ samples is labeled. Next, the DNN is retrained on all the labeled samples and the linear graphs are updated for each class. More details about GALAXY and evaluations on non-RF datasets are described in~\cite{galaxy}.


\begin{figure*}[t!!!]
    \centering
    \includegraphics[width=0.8\linewidth]{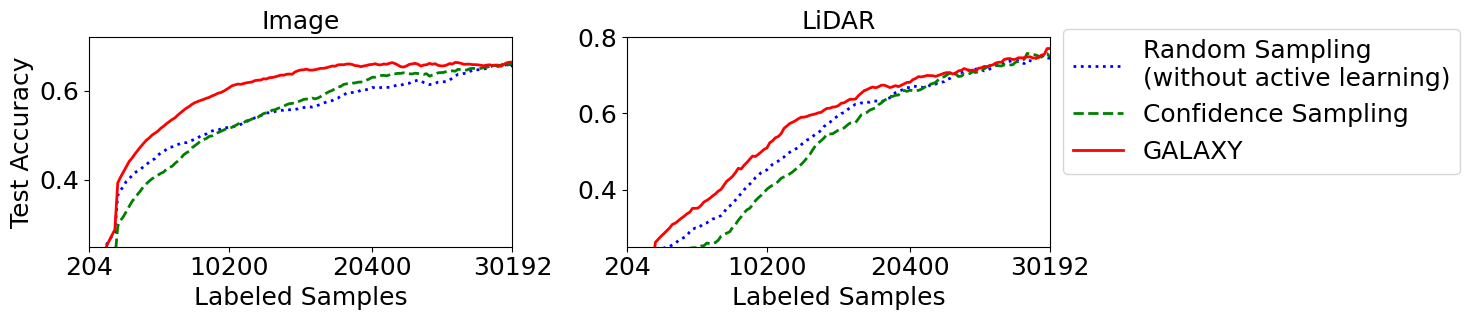}
    \caption{Average test set accuracy measured in each active learning iteration for two modalities of image and LiDAR, each with three different algorithms of random sampling, confidence sampling, and GALAXY.}
    \label{fig:test-acc}
\end{figure*}

\begin{figure*}[t!!!]
    \centering
    \includegraphics[width=0.8\linewidth]{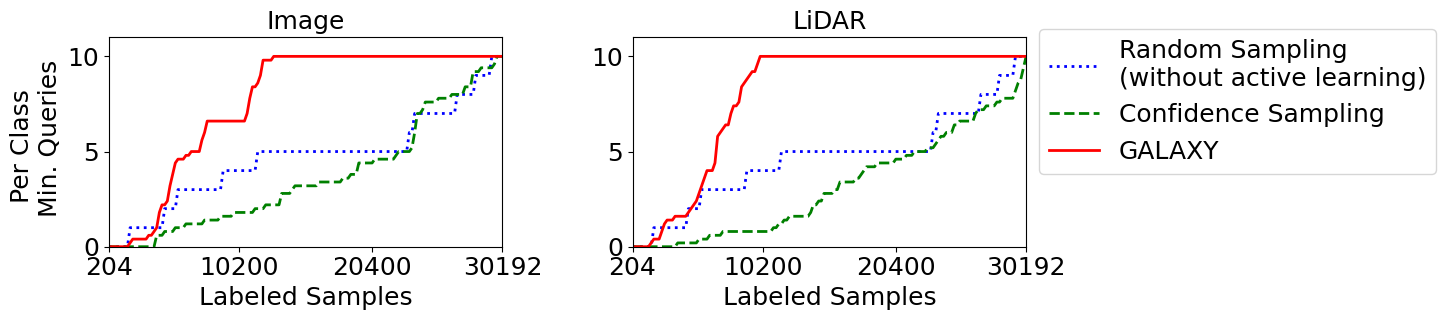}
    \caption{Average per-class minimum queries on the training set in each active learning iteration for two modalities of image and LiDAR, each with three different algorithms of random sampling, confidence sampling, and GALAXY.}
    \label{fig:min-queries}
\end{figure*}


\section{Evaluations}\label{sec:eval}
In this section, we show test accuracy results on two modalities of FLASH dataset for mmWave beam selection with and without active learning. 

\subsection{Evaluation Setup}
As the lowest class population is 20 (see Fig.~\ref{fig:population-per-class}), we shuffle the dataset and pick 10 samples from each class to compose the test set, and keep the rest (30371 samples) for training. In this way, we maintain a class-balanced test set where all classes contribute equally to test accuracy, and an imbalanced training set, which is the real-world case for any in-the-wild RF training dataset collection. We use GALAXY~\cite{galaxy} as the state-of-the-art active learning algorithm for imbalanced datasets. To have a point-to-point comparison with classical learning in each and every iteration, we use \emph{random sampling} to \emph{randomly} select samples to be labeled~\cite{galaxy}. We also compare GALAXY with confidence sampling that is a popular uncertainty-based active learning algorithm~\cite{holzmuller2022framework}.

For each modality, we use different DNNs. For image modality we resize the inputs to dimensions of (3, 90, 160) and use standard ResNet18 with 11.1M parameters from \texttt{torchvision} package. For LiDAR modality with inputs of size (20, 20, 20), we use a custom convolutional DNN with residual blocks totally with 1.1M parameters, that is used in~\cite{flash}. Following the pool-based active learning steps described in Section~\ref{sec:categories}, we set the active learning batch size to 204 which means in each iteration 204 samples are selected and queried for labels and are added to the current labeled training set. As explained before, the strategy of selecting the new batch from the unlabeled pool of samples depends on the active learning algorithm. We train each DNN with its corresponding training set modality of that iteration for several epochs, until it is fully trained. At the end of each iteration we measure test accuracy and per-class minimum queries, which is a good metric for class-diversity. We repeat the end-to-end training/test experiments 5 times for each active learning algorithm in each modality, and between the 5 runs, we shuffle and re-partition our training and test sets. For each modality and each algorithm, we report the average of test accuracies at each iteration over the 5 runs. We also report standard errors (calculated as standard deviation of accuracies divided by square root of number of runs) of test accuracies at each iteration averaged over all iterations.

\subsection{Numerical Results}
\subsubsection{Test Accuracy}
For each modality and each algorithm, we smoothen the curves of average test accuracies using an averaging window of size 10, for less spiky illustration. In Fig.~\ref{fig:test-acc}, we show average test accuracy versus the number of labeled samples in each iteration. We observe that for both modalities GALAXY algorithm plateaus in earlier iterations with fewer labeled samples.
For image modality, we report average standard errors of 1.0\%, 1.2\%, and 1.0\% for random sampling, confidence sampling, and GALAXY, respectively. This standard error for each algorithm is caused by different training and test set partition selections, as well as different seeds for the initial random batch selection among all runs.

In image modality, we observe that random sampling, confidence sampling, and GALAXY achieve an example accuracy of 60\% with 19176, 17136, and 9588 labeled samples, respectively. This means that GALAXY is able to achieve the same accuracy of 60\% with 50\% and 40\% fewer labeled samples compared to random sampling and confidence sampling, respectively. In the same modality, GALAXY can achieve 63\% accuracy with 42\% and 41\% fewer labeled samples compared to random sampling and confidence sampling, respectively.

For LiDAR modality, we plot the average accuracies and report average standard errors of 1.6\%, 1.8\%, and 1.7\% for random sampling, confidence sampling, and GALAXY, respectively. We observe that GALAXY is able to achieve an example accuracy of 71\% with 11\% and 10\% fewer labeled samples compared to random sampling and confidence sampling, respectively.

\subsubsection{per-Class Minimum Queries}
We showed that GALAXY excels the performance of random sampling and confidence sampling by achieving the same accuracy with fewer number of labeled training samples (a.k.a queries). Here we show that GALAXY achieves this by considering the predicted class for selecting the samples to be queried, and selects a class-diverse batch. Fig.~\ref{fig:min-queries} shows average minimum queries per class for image and LiDAR, with three different algorithms of random sampling, confidence sampling, and GALAXY. We recall that the smallest class has 20 members (see Fig.~\ref{fig:population-per-class}) out of which 10 are partitioned as the test set. Therefore, the smallest class in the training set has 10 members. In Fig.~\ref{fig:min-queries}, we observe that GALAXY reaches per-class minimum queries of 10 much earlier than random and confidence sampling, which shows that the smallest class is completely queried in earlier iterations. We observe that in the image modality, GALAXY reaches per-class minimum queries of 10 at 12648 labeled samples, while random and confidence sampling get to the point of completely sampling the smallest class when 29376 and 29784 samples are queried, respectively. This shows that GALAXY fully queries the smallest class in the training set in 56\% and 57\% fewer queries compared to those of random and confidence sampling, respectively. Similarly, in LiDAR modality, GALAXY fully queries the smallest class in 66\% and 67\% fewer queries compared to random sampling and confidence sampling, respectively.


\section{Future Directions}\label{sec:future}
\noindent
$\bullet$ {\bf Digital Twins of Wireless Networks:} A digital twin is a virtual model of a real world environment that is designed to study the properties of the real world without risking damage to life or property in it. In wireless communications, digital twins are used for modeling RF propagation patterns, designing network architectures, and optimizing PHY protocols. With a high-fidelity digital twin, the emulation outputs are analogous to real-world observations. As a result, recent work suggest using digital twins instead of running measurement campaigns and generating labels for machine learning tasks. While using digital twins significantly reduces human effort and equipment cost, running a high fidelity digital twin for labeling a dataset often requires intensive computation and is time consuming. On the other hand, active learning enables training DNNs with a reduced number of labeled training samples. As a result, by pairing the digital twins with active learning, a framework can be self-sufficient by optimally generating the labels in the digital twin. 

\noindent
$\bullet$ {\bf Active Learning in Quantum Communication:} 
Quantum communication is used for transmitting highly sensitive data due to the entanglement process, where eavesdropping leaves a trace, as measuring the state of one qubit affects the state of another qubit that is entangled with it. 
Hence, while designing a deep-learning-based receiver for quantum communications, labeling qubits that involves measuring their states is expensive and consequential. In this case, active learning can help train an equally robust DNN with fewer labeled samples (qubits).

\noindent
$\bullet$ {\bf Active Learning for Preserving Privacy in Open Radio Access Network (O-RAN)}: Active learning can be specially helpful when a training set is collected using an O-RAN system, as the data and labels have high privacy in O-RAN systems. Active learning can help reduce the number of required labeled samples, and hence preserve user's privacy as much as possible. In this case, the network operators can start with a few iterations of random sampling until there is one labeled sample from each class. Then, a more sophisticated sampling algorithm could be used for more guided sampling and selecting the most informative samples.

\noindent $\bullet$ \textbf{Optimizing for Training Computational Cost Besides Labeling Cost}: As established in this paper, active learning aims to reduce the labeling cost through adaptively and iteratively selecting the most informative unlabeled samples to be queried for labels. Apart from this, continual and life-long learning algorithms that are originally designed to address the \emph{catastrophic forgetting} problem, can also be used for optimizing training computational cost. This is achieved by preserving the previous knowledge and fine-tuning the DNN using the new data, instead of re-training it again on old and new data from scratch. The two categories of methods are mutually exclusive and can be jointly applied to deep learning based PHY applications to improve both labeling and training computational costs.

\section{Conclusion}\label{sec:conclusion}
In this paper, we introduced active learning for deep learning applications in wireless communications. We described different categories of active learning algorithms and mapped them to different PHY deep learning applications. Next, we discussed a case study of mmWave beam selection as an example of active learning for extremely class-imbalanced datasets, that is the case for many RF datasets that are collected in the wild. We investigated how active learning reduces labeling overhead for two different modalities in the dataset, and showed that using active learning we can achieve the same accuracy as the classical training with up to 50\% fewer labeled samples. We further showed future directions for using active learning in wireless communications.


\section*{Acknowledgement}
This work is supported by AI Institute for Future Edge Networks and Distributed Intelligence (AI-EDGE) NSF award \#2112471.


\bibliographystyle{ieeetr} 
\bibliography{ref} 

\section*{Biographies}

\noindent \textbf{Nasim Soltani} (soltani.n@northeastern.edu) is a PhD student at Northeastern University. Her research interest is broadly machine learning applications for the physical layer of wireless systems.\\

\noindent \textbf{Jifan Zhang} (jifan@cs.wisc.edu) is a PhD student at the computer science department at the University of Wisconsin-Madison. His research focuses on label-efficient learning and human-in-the-loop learning on large scale deep learning systems.
\\

\noindent \textbf{Batool Salehi} (salehihikouei.b@northeastern.edu) is pursuing a Ph.D. degree in computer engineering at Northeastern University. Her current research focuses on mmWave beamforming, IoT, and the application of ML in the domain of wireless communication.
\\

\noindent \textbf{Debashri Roy} (debashri.roy@uta.edu) is currently an Assistant Professor in University of Texas at Arlington. Her research interests are in the areas of machine learning and wireless communication.
\\

\noindent \textbf{Robert Nowak} (rdnowak@wisc.edu) currently holds the Keith and Jane Nosbusch Professorship in Electrical and Computer Engineering at the University of Wisconsin-Madison. His research focuses on machine learning, optimization, signal processing and statistics.\\

\noindent \textbf{Kaushik Chowdhury} (krc@ece.neu.edu) is professor at Northeastern University. His current research interests involve systems aspects of networked robotics, machine learning for agile spectrum sensing, and large-scale experimental deployment of emerging wireless technologies.

\end{document}